# DeFi TrustBoost: Blockchain and AI for Trustworthy Decentralized Financial Decisions

*Dr. Swati Sachan*
*Artificial Intelligence in Finance*
University of Liverpool
Financial Technology
Chatham St, Liverpool, L69 7ZH
United Kingdom
**swati.sachan@liverpool.ac.uk**

*Prof. Dale S. Fickett*
*Entrepreneurship Professor*
Robins School of Business
University of Richmond
Richmond, Virginia, VA 23173
United States
**d.fickett@richmond.edu**

# DeFi TrustBoost: Blockchain and AI for Trustworthy Decentralized Financial Decisions

A B S T R A C T

This research introduces the Decentralized Finance (DeFi) TrustBoost Framework, which combines blockchain technology and Explainable AI to address challenges faced by lenders underwriting small business loan applications from low-wealth households. The framework is designed with a strong emphasis on fulfilling four crucial requirements of blockchain and AI systems: confidentiality, compliance with data protection laws, resistance to adversarial attacks, and compliance with regulatory audits. It presents a technique for tamper-proof auditing of automated AI decisions and a strategy for on-chain (inside-blockchain) and off-chain data storage to facilitate collaboration within and across financial organizations.

## 1. Introduction

Low-wealth households encounter difficulties in raising funds to launch a small business venture. These endeavors demand considerable capital, which could be challenging to acquire through conventional means (Alvarez & Barney, 2014). Traditional financial institutions, such as retail banks, have rigorous underwriting criteria which complicate loan approval for individuals lacking sufficient financial assets and strong credit history (Sachan, Yang, Xu, Benavides, & Li, 2020).

Individuals facing challenges in securing loans seek guidance on their probability of success through an advisor or automated AI (Artificial Intelligence) systems without engaging in a formal decision-making process. These underserved individuals are vulnerable to biases and discrimination associated with gender, ethnicity, and other societal factors. Their underrepresentation presents a shortcoming in many AI training datasets (Kordzadeh & Ghasemaghaei, 2022). Collaboration among financial experts from various organizations can enrich these datasets. This research introduces the Decentralised Finance (DeFi) TrustBoost framework, which integrates blockchain technology with an explainable deep-neural-network (x-DNN) to develop a trustworthy AI advisory system to address the following challenges:

a) *Decentralised sharing mechanism*: A decentralized data-sharing approach can promote joint data-sharing incentives by resolving issues related to personal identity concerns and centralized data access control (Yang, Garg, Huang, & Kang, 2021).

b) *High-stake financial decisions by explainable AI*: Stakeholders prioritize explainability in AI decisions for critical situations involving human lives or significant assets, despite the availability of high-performance AI algorithms such as Deep-Neural-Networks (DNNs) (Ozbayoglu, Gudelek, & Sezer, 2020). DNNs are widely applied in finance, such as algorithmic trading (Goutte, Le, Liu, & von Mettenheim, 2023) (Liu, Li, Li, Zhu, & Yao, 2021), risk assessment (Zhu, Zhang, Wu, Li, & Li, 2022), fraud detection (Xiuguo & Shengyong, 2022) (Roy, et al., 2018), and portfolio management (Betancourt & Chen, 2021) (Lucarelli & Borrotti, 2020). It has faced criticism due to its inability to provide human-level reasoning behind decisions. To address this, explainable AI (XAI) researchers have developed model-agnostic and model-specific techniques to understand the reasoning behind decisions (Montavon, Samek, & Müller, 2018).

c) *Defence against adversarial attack*: The centralized AI solutions are susceptible to adversarial attacks by malicious actors manipulating the algorithm's environment (Nassar, Salah, ur Rehman, & Svetinovic, 2020). Such attacks could involve tampering with model parameters, decision boundary alterations, introducing noisy data samples, or modifying the rationale behind loan application acceptances and rejections. These assaults can lead to incorrect decisions resulting in substantial financial losses, devastating effects on the lives of underserved individuals, and potential legal ramifications.

## 2. DeFi TrustBoost Framework

The DeFi TrustBoost architecture has two main components: the integrated operational workflow for blockchain and x-DNN model based on one-dimensional Convolutional Neural Networks (1D-CNN) for human-in-the-loop decision-making. It focuses on four crucial requirements: confidentiality, compliance with data protection laws (Union, 2016) (Finck, 2019), resistance to adversarial attacks, and the ability to withstand regulatory audits, shown in Figure 1. Table 1 provides definitions for the four main actors and DeFi TrustBoost framework illustrated in Figure 2.

**Design Requirement of DeFi TrustBoost Framework**

| K1: Confidentiality | K2: Compliance with Data Protection Laws |
|---|---|
| • Secured storage of the identity of actors<br>• Secured storage of data<br>• Decentralised record of consents, data accessed, and decision by XAI system | • Consent to access the personal data<br>• Europe: Adherence to GDPR's Article 17, "Right to be forgotten," to ensure personal data is not retained without consent<br>• Europe: Compliance with GDPR's Article 22, "Right to explanation," to provide transparency for decisions made by AI algorithms<br>• US Data Privacy: Gramm-Leach-Bliley Act; Fair Credit Reporting Act |
| K3: Adversarial Attack Resistance | K4: Regulatory Audits |
| • Robust model architecture to withstand adversarial attacks<br>• Regular monitoring and updating of the XAI system | • Maintaining detailed records of algorithmic parameters and architecture<br>• Implementing internal audit processes and cooperating with external audits<br>• Ensuring transparency in algorithmic operations and decision-making |

Fig. 1. Requirement Analysis

### 2.1 Web-Application for Expert Knowledge Elicitation

All organizations engaged in data collection through knowledge elicitation from financial experts utilize a user-friendly web application (first layer: Figure 2 (A)). It facilitates interaction with experts to quantify heuristic knowledge through structured queries. The experts can monitor and manage their data usage activities and consent.

### 2.2 Interoperability: Blockchain, Web-API & Other Services

The API server (second layer: Figure 2 (A)) enables interoperability between services: the web application, blockchain platform (third layer: Figure 2 (A)), off-chain cloud-based data storage, and secured cloud-based cryptographic-key management.

| \multicolumn{2}{c}{**Table 1: Actors within Organisations**} |
|---|---|
| \multicolumn{2}{c}{**Definition of organisation**} |

Organisations, denoted by $o$ ($o \in \{1, ..., O\}$), provide infrastructure and resources for the AI-based decision-making system. They enforce policies for ethical AI, ensure compliance with regulations for shared data usage in loan application processing, and bear legal responsibility for the system's actions. In Figure 2, $o \neq o'$ and $o, o' \in \{1, ..., O\}$, represents two distinct organisations.

| Actors in organisation | Role of actors |
|---|---|
| Human experts (Underwriters and Lending Experts) | Data subjects, denoted by $h$ ($h \in \{1, ..., H\}$), are human experts who:<br>1. Contribute knowledge on structured questions or statements through a web application<br>2. Serve as finance professionals within an organisation, responsible for manual decision-making on loan applications when the AI model yields low-confidence outcomes |
| Developer (or Data Scientist) | Developers or data scientists, denoted by $u$ ($u \in \{1, ..., U\}$):<br>1. Manage off-chain and on-chain (blockchain platform) data storage<br>2. Retrieve off-chain data for the development of an XAI-based decision-support system for advisory<br>3. Retrieve on-chain data for internal audit of consents granted by data subjects and explanations and decisions by the XAI system |
| Audit Regulator | Audit regulators, denoted by $z$ ($z \in \{1, ..., Z\}$):<br>1. Assess the quality and consistency of a loan application decision and its explanations generated by the XAI model<br>2. Investigate potential breaches of consents granted by data subjects concerning their shared expertise and personal identity |
| Customer | Customers, denoted by $x, (x \in \{1, ..., X\})$:<br>3 Small business owners for whom the XAI system provides decisions<br>4 Susceptible to potential risks from AI decision-making systems and manual underwriting of experts attributable to biases caused by under-representation in training data and inaccuracies arising from the complex loan applications, respectively (Sachan, Yang, Xu, Benavides, & Li, 2020) (Sachan S., 2022). |

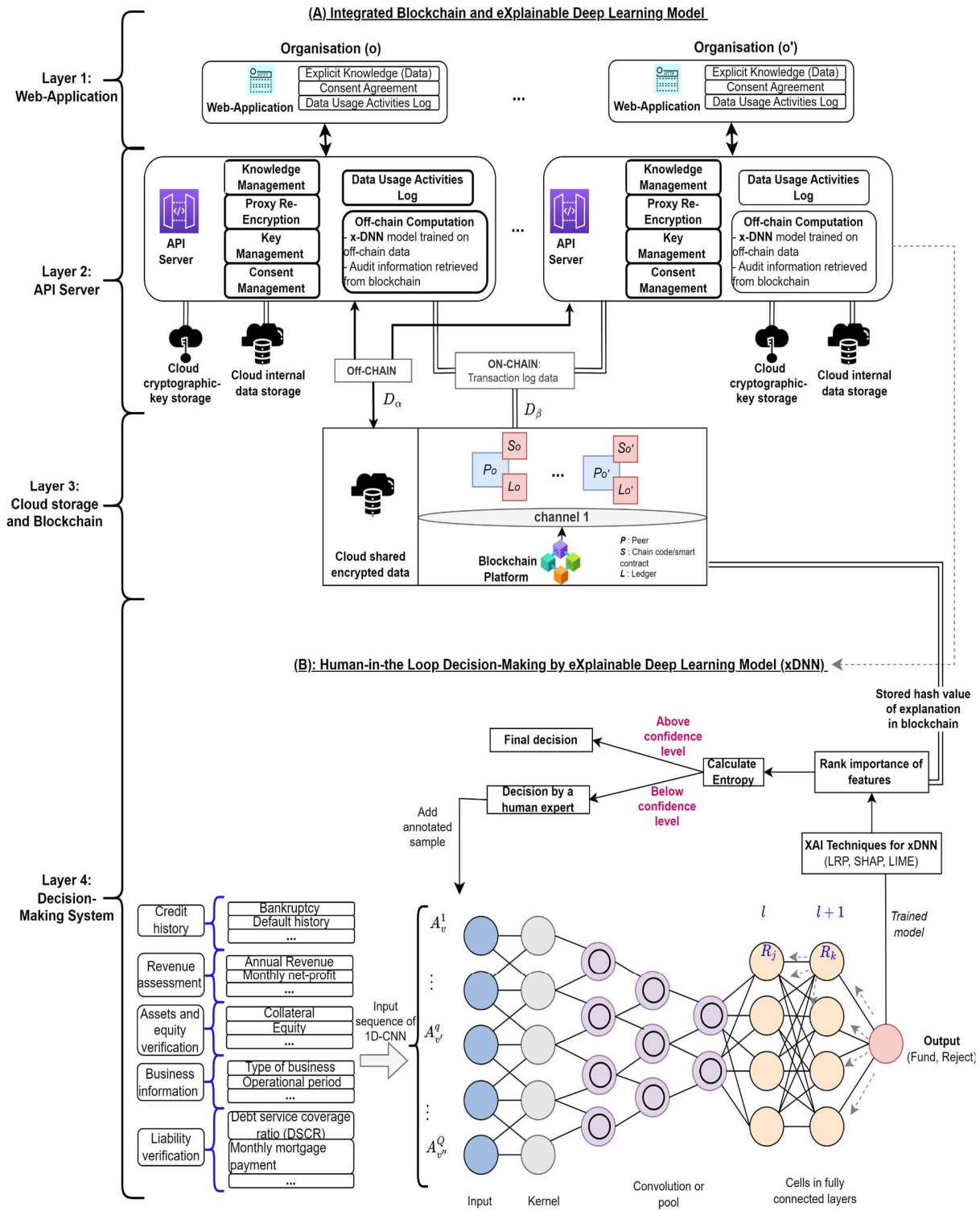

**Fig 2.** DeFi Trustboost Architecture to Integrate Blockchain and Explainable AI

## 2.3 Off-Chain and On-Chain Data Storage

Off-chain data storage, such as a secured cloud, refers to keeping data outside the blockchain. On-chain data storage involves maintaining data within the blockchain, like Ethereum or Hyperledger. Decentralized finance using blockchain technology faces two obstacles:

- Blockchain's immutability clashes with global data protection regulations (see Table 1), restricting the permanent storage of personal or sensitive information.

- Blockchain is designed to store simple transaction records, not large data file systems. For example, if the storage cost of a kilobyte in Ethereum costs 0.032 ETH or $50.93, then a 2000-byte loan application form in Ethereum costs $101.86, while a 100-bytes payment transaction costs $5.093. This cost is much higher than cloud storage alternatives like Amazon Blob charges $1.8 for the first 50 terabytes.

A practical blockchain implementation for decentralized finance requires optimization of on-chain data storage costs to mitigate high fees and computational expenses associated with processing complex on-chain information in smart contracts (SC). Striking the right balance in on-chain data distribution is crucial, as excessive transparency can lead to overexposure. In contrast, too little data may lead to a lack of trust.

Table 2 demonstrates the alignment of off-chain and on-chain data storage strategies with the framework's design requirements. Data gathered from multiple organizations are stored as encrypted files in a shared cloud repository. Encrypted data, such as an XAI model, visual or text explanations of loan decisions, and encrypted training and test data files, is maintained in each organization's private cloud repository. Direct storage of this data on the blockchain is infeasible due to limited block size, the right to deletion imposed by data protection laws, and high fees. Therefore, only the corresponding hash values of these files are stored within the blockchain to enable auditing for data tampering and detection of adversarial attacks.

| Table 2: Off-Chain and On-Chain Data Storage Strategy | | |
|---|---|---|
| Storage | Data Type | Mapping with Design Requirements |
| **Off-Chain or Secured Cloud** ($D_\alpha$) | Personal identity: experts contributing data samples and customers processed by an XAI system | K1, K2 |
| | Encrypted training & validation data utilized in XAI system deployment | K2, K3 |
| | Encrypted XAI model configuration: hyperparameters and parameters | K3 |
| | An encrypted explanation for a given decision by the XAI model | K3, K4 |
| **On-Chain or Blockchain** ($D_\beta$) | Consents by experts | K2, K4 |
| | Hashed personal ID/credential of an individual on behalf of the organization | K1, K3 |
| | Hashed XAI model configuration | K3 |
| | Hashed explanation for a given decision by the XAI model | K3, K4 |

**2.4 Blockchain-based Tamper-Proof Auditing for XAI Systems and Data Consent**

The immutability of a blockchain ledger enables trust among multiple organizations sharing expertise to improve the data representation of underserved individuals. An interpretable and adversarial attack-resistant XAI system extends the trust in customers, users, and regulators, who confidently accept automated decisions.

XAI systems interpret decisions by model-agnostic and model-specific techniques in visual or textual formats. The process of auditing the tampering of the XAI system's dynamics is illustrated in Figure 3. It displays an image (PNG) and a text file of a decision explanation for the $x^{th}$ customer's loan application. These files are encrypted and stored in a cloud database, while their hash values are stored in the blockchain. The one-way hash function cannot revert the hash value to the original data, unlike encrypted data, which is decryptable. A hash is a fingerprint of the information without disclosing the actual information. This framework uses proxy-encryption (Manzoor, Braeken, Kanhere, Ylianttila, & Liyanage, 2021) and the SHA256 algorithm for encryption and hashing, respectively. The storage of explanation $E$ for $x^{th}$ customer in cloud data storage denoted by $D_\alpha^x$ and blockchain denoted by $D_\beta^x$ can be represented as:

$$D(E, x) = \begin{cases} D_\alpha^x = \left((ID^x, DTM, H_E^x), Enc_E^x\right) \\ D_\beta^x = (ID^x, DTM, H_E^x) \end{cases} \quad (1)$$

where, $ID^x$ is the customer identification, $H_E^x$ is the hash value of the decision explanation, $Enc_E^x$ is the encrypted file of the explanation, and $DTM$ signifies the date and time.

The tampering state is denoted by $\tau$, where $\tau = 0$ represents no tampering and $\tau = 1$ indicates tampering. To verify the consistency of a decision given to a customer $x^{th}$, an auditor first retrieves off-chain data $\left((ID^x, DTM, H_E^x), Enc_E^x\right)$ using the customer's identification. Then, an SC is deployed to retrieve the triple $(ID^x, DTM, H_E^x)$ stored in the blockchain. The non-existence of triple indicates a discrepancy between off-chain and on-chain data, suggesting tampering $\tau = 0$. If the triple is found, the hash value of the decrypted explanation is compared for further verification. The tampering conditions are:

$$\tau_{XAI} = \begin{cases} \tau = 1, if \begin{cases} (ID^x, DTM, H_E^x) \not\exists\, D_\beta^x \\ H(DeCr_E^x) \neq H_E^x \end{cases} \\ \tau = 0, otherwise \end{cases} \quad (2)$$

Auditing data consent shared among multiple organizations resembles auditing an XAI system. Expression (3) represents the complete data consent state for expert $h$ of an organization $o$, which includes the consent for data acquisition ($CAQ$), consent to withdraw data ($CW$), and consent to access ($CA$). All organizations store the consent state of contributing experts across all organizations in their private cloud database. SC then pushes the hash value of consent to the blockchain.

$$\mathbb{C}_{h,o} = \begin{cases} CAQ_{h,o} = \begin{cases} approved \\ invalid \\ rejected \\ awaiting \end{cases} \\ CW_{h,o} = \begin{cases} requested \\ invalid \\ not\ requested \end{cases} \\ CA_{h,o} = \begin{cases} agreed\ \&\ valid \\ invalid \\ not\ agreed \\ awaiting \end{cases} \end{cases} \quad (3)$$

Expression (4) indicates consent tampering when the hash values of states from two distinct organizations do not exhibit strict equality.

$$\tau_\mathbb{C} = \begin{cases} \tau = 1, H(\mathbb{C}_{h,o}) \neq H(\mathbb{C}_{h,o'}) \\ \tau = 0, H(\mathbb{C}_{h,o}) = H(\mathbb{C}_{h,o'}) \end{cases} \quad (4)$$

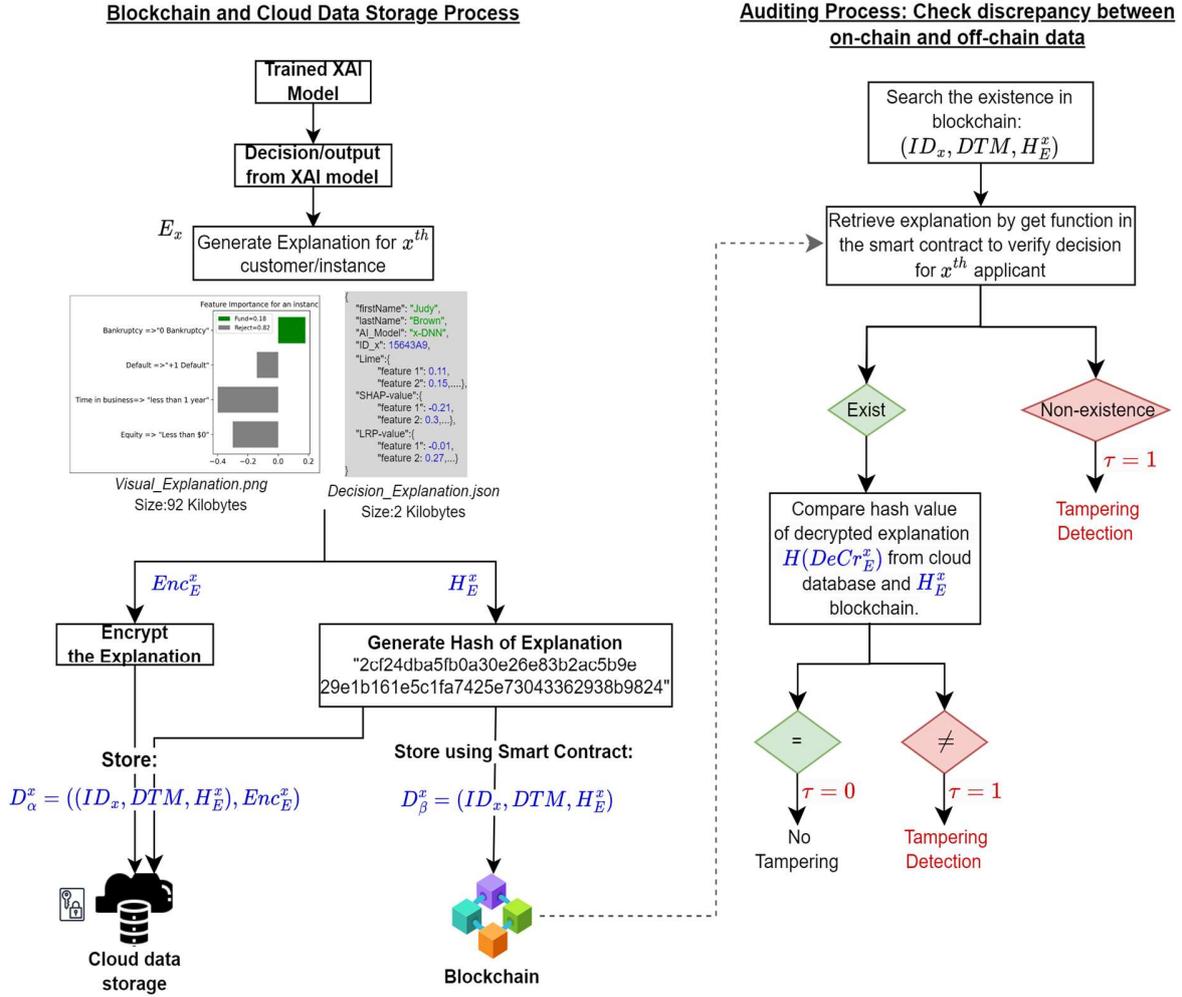

**Fig 3.** Blockchain-based tamper-proof auditing of the XAI system's decision explanation

## 2.5 Human-in-the-Loop Learning by XAI

The choice of an AI algorithm for decision support depends on the use case and stakeholders within an organization. DNNs are recognized for their high accuracy but criticized for their opacity in revealing the reasoning behind a decision, which is imperative for automated advice or decisions on small business loans. The explainability of DNNs for structured data, such as loan information, is limited compared to unstructured data, images, texts, and speech. Commonly used model-agnostic techniques can explain decisions made by black-box AI algorithms trained on structured data: Local Interpretable Model-Agnostic Explanations (LIME) (Ribeiro, Singh, & Guestrin, 2016) and SHapley Additive exPlanations (SHAP) (Lundberg & Lee, 2017).

CNNs are a subtype of DNN architecture primarily used for image recognition. For unstructured data, CNNs benefit from the inherent visual structure of input data to highlight the portion of the image which contributes towards a decision. The model-agnostic approaches can help explain CNNs, a more accurate model-specific technique called Layer-wise Relevance Propagation (LRP) (Bach, et al., 2015).

This framework showcases the accuracy, visual explainability, and tamper-proof auditing process for a 1D-CNN designed for small business loan advisory, illustrated in fourth layer of Figure 2 (B). An input feature in 1D-CNN is denoted $A_v^q$ where $q \in \{1, ..., Q\}$ is number of features and $v \in \{1, ..., V\}$ is cardinality of transformed feature.

LRP, SHAP, and LIME, shown in Table 3, are used in conjunction to visualize and interpret the network's loan application decisions $\{Fund, Reject\}$ based on a 1D-CNN model trained on small business loan data. The 1D-CNN aims to assist financial advisors or loan underwriters in daily tasks, providing reliable decisions for underserved individuals. The system actively learns from experts who provide manual decisions for cases with low-confidence outcomes, which are subsequently used as annotated samples to retrain the system.

Active learning of the 1D-CNN involves incorporating feedback from human experts to fine-tune the model for uncertain decisions. Entropy estimates the uncertainty in a decision; low entropy indicates high confidence for a given prediction and vice versa (Wan, et al., 2020). The entropy score for the prediction of each instance:

$$\emptyset(x) = \frac{-\sum_{\theta \in P(\theta)} P(\theta|x) \log_2 P(\theta|x)}{\log_2(N)}, \emptyset(x) = [0,1] \tag{5}$$

where, $P(\theta|x)$ is the predicted outcome for an instance $x$ each decision $\theta$ and $N$ is the number of decisions. The predicted data samples are arranged based on the entropy score.

| Table 3: Techniques to Generate Decision Explanation by XAI Model Based on 1D-CNN | | | |
|---|---|---|---|
| **Technique Type** | | **Technique** | **Visual Explanation** |
| **Model-Specific (Ante-hoc)** | LRP | LRP operates by recursively propagating relevance scores from the output layer to the input layer. It assumes that total relevance, such as the activation strength of an output node for a specific class, is conserved within each layer. The relevance ($R_k$) of a neuron $k$ in layer $l + 1$ is the sum of the relevance of all nodes $j$ in layer $l$ contributing to the neuron $k$, shown in Figure 2 (B). There are four types of LRP: | Heatmap: Colour scheme to reflect the positive and negative contributions towards the final decision (target class). |

|  |  | (a) LRP-0: baseline methods which redistribute relevance in proportion to the contributions of each input to the neuron activation<br>(b) LRP-$\epsilon$: absorbs some relevance of weak contributing neuron<br>(c) LRP-$\gamma$: reduces the noise and reduces negative contribution over positive<br>(d) LRP-$\alpha\beta$: treat positive and negative contributions asymmetrically. |  |
|---|---|---|---|
| **Model-Agnostic (Post-hoc)** | LIME | It utilizes interpretable algorithms, such as linear regression and decision trees, as a surrogate model, which approximates the behavior of the original non-interpretable algorithm in the proximity of a specific data sample. | Bar Plot: normalized importance of features |
|  | SHAP | It is based on coalitional game theory, wherein input features act as players, and the predicted value signifies the total game reward. Shapley values are computed to determine each feature's marginal contribution by averaging the differences in the model's output when the feature is added or removed across all possible feature subsets. | Bidirectional Bar Plot: Display positive or negative contributions to the final decision (target class). |

## 3. Results on Implementation of DeFi TrustBoost

### 3.1 Blockchain Performance

The study used two open-source blockchain networks: Ethereum, a public platform, and Hyperledger, a private and permissioned one. The test environment configuration and technologies for actual implementation is shown in Table 4. Blockchain chronologically stores the hashes of explanations and data subject consents. Its adoption in decentralized financial applications hinges on operational efficiency, gauged through latency and throughput. Latency refers to the delay in block addition, while throughput is the count of valid transactions processed by the blockchain. Figure 4 demonstrates the performance in terms of transaction latency, latency by a varying number of nodes, and throughput for a dataset of 800 transactions across four organizations in different nodes. The presented results reveal that Hyperledger outperforms Ethereum. It can serve as a viable tool for archiving consents and decisions derived from Explainable AI (XAI) models, facilitating auditing. However, as the number of nodes and transactions increases, there is a noticeable increase in latency and a corresponding decline in throughput. It suggests that blockchain may not be the best solution for real currency transactions due to potential risks such as double-spending.

The efficiency of the audit process was assessed by introducing random alterations in 2% to 20% of the files stored in off-chain mediums. The goal was to ensure that the recomputed hash of off-chain storage mediums matched their respective hash permanently stored in a blockchain platform (on-chain) to pass the automated audit. The completion time of the audit process grows with the increase in the number of tampered files in both Ethereum and Hyperledger Fabric.

| Table 4: Test Environment | |
|---|---|
| **Component** | **Description** |
| Organisations ($o$) | 4 organisations |
| Experts ($h$) | 12 data subjects: 3 experts from each organization |
| Test Machines | 4 machines running Ubuntu 19.04, 24 Core CPU, and 64 GB RAM |
| API Server | FastAPI: a framework for REST-API |
| Blockchain Network (On-chain Data Storage) | Hyperledger Fabric Version 1.4 and Ethereum |
| Proxy-encryption key and Identity Storage | Azure Key Vault |
| Off-chain Data Storage | Azure Blob Storage |
| Batch Size | |

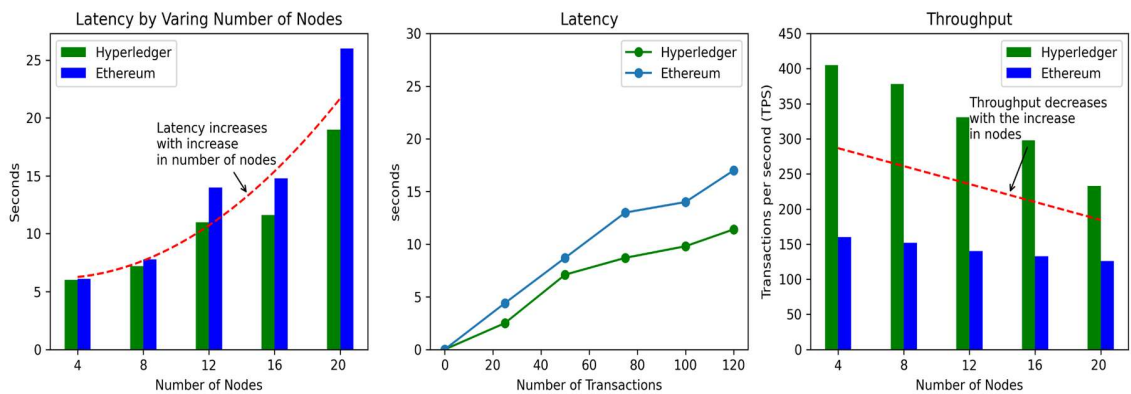

**Fig 4.** Ethereum and Hyperledger Platform Performance

## 3.2 Decision Explanation and Human-in-the-Loop Involvement

The dataset contained 1888 samples, with 45.52% labeled as funded and 54.48% as rejected. It consisted of 18 attributes, transformed into 88 numerical features using a one-hot encoder. The optimal architecture for the 1D-CNN model was determined through Bayesian optimization, as presented in Table 5.

Figures 5 and 6 provide visual explanations of probabilistic loan application decisions interpreted by LRP, LIME, and SHAP. The LRP heatmap's diagonal hatch pattern represents attributes of high importance (values > 0.50), while its dimensionality matches the number of input nodes activated by the 1D-CNN. Among these, LRP-$\gamma$, LRP-$\alpha\beta$, and SHAP techniques provided more accurate explanations. These comprehensive visual explanations can be effectively utilized by underwriters or loan advisors to advise on loan acceptability and define requirements for loan acquisition. To ensure data integrity and transparency, the hash value of the PNG files containing these visual explanations is securely stored on-chain within a blockchain network. This approach allows for an efficient audit of past decisions and helps identify any potential data tampering. The model's hyperparameters and trained parameters are also stored in the blockchain to enable regular checks to mitigate potential adversarial attacks.

| Table 5: Trained Hyperparameters and Architecture of 1D-CNN | | |
|---|---|---|
| **Layer and Other Parameters** | **Description** | **Parameter** |
| *Number of input layers* | Sequence size (height × width) for the 1D-CNN | 88×1 |
| *Number of convolution and pooling layers* | Convolution layers is followed by a maximum pooling layer. Maximum pool selects the largest element within the region covered by the filter/kernel. | Number of convolution and pooling layer layers = 3<br>Kernel size in three layers = {5, 5, 2}<br>Number of kernels in three layers = {50, 50, 60}<br>Activation function = $LReLU$<br>Number of strides in three layers = {1, 2, 2}<br>$L^2$ regularization strength in layers = 0.01 |
| *Fully Connected Layer* | Output from the previous maximum pooling is flattened to form two fully connected or dense layers | Number of fully connected layers = 2<br>Dropout rate = 0.10<br>Activation function = $LReLU$<br>$L^2$ regularisation strength = 0.01 |
| *Output layer* | Predicts outcome for an applicant | Activation function = $SoftMax$<br>$L^2$ regularization strength = 0.01 |
| *Batch size* | Number of training samples in each iteration | 100 |

| Learning rate | The step size at each iteration while moving toward a minimum of a loss function | 0.0001 |

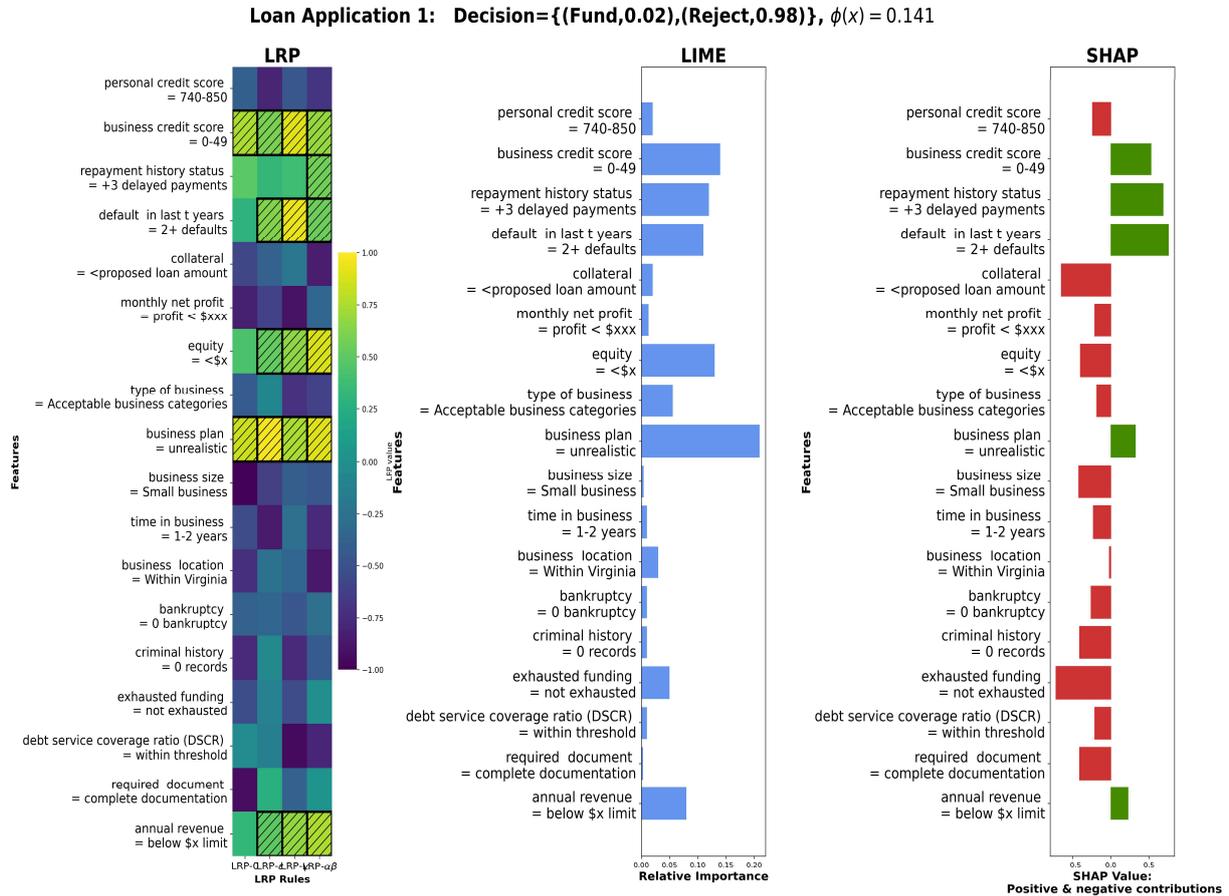

**Fig 5.** The rationale for rejecting Loan Application 2 with a 0.98 probability, supported by a confidence level within the appropriate threshold ranging from $0 \leq \emptyset(x) = 0.141 \leq 0.80$

The system actively learns from financial experts to incorporate feedback from financial experts who manually decide on cases with low-confidence outcomes. Figure 7 reports the AUC metrics from a 5-fold cross-validation across six training iterations of the 1D-CNN model, each iteration incorporating an additional 150 expert-annotated instances. It demonstrates an improvement in performance over time: the overall AUC value has increased from 0.74 in the initial baseline iteration, which did not include any expert-annotated data, to 0.92 by the sixth iteration.

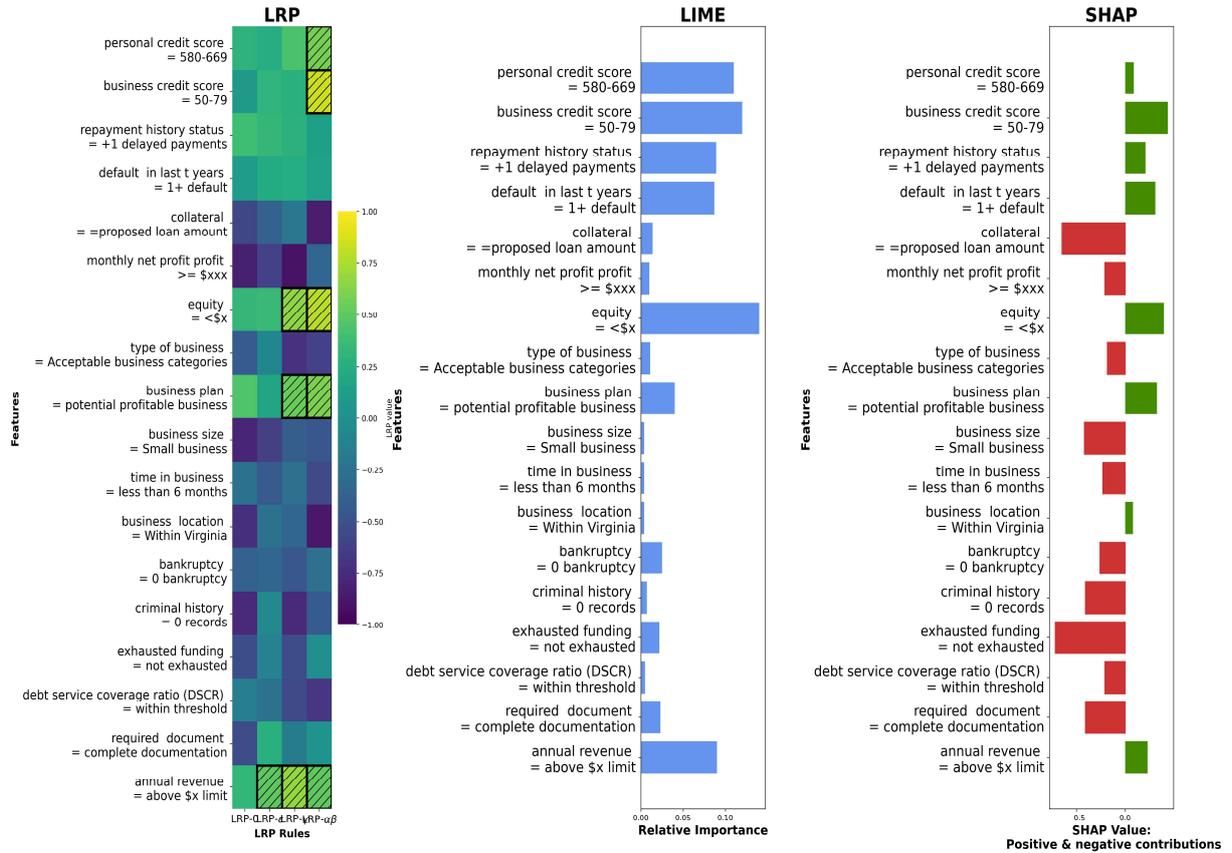

**Fig 6.** Reasoning for a human expert's manual evaluation of Loan Application 2, due to the low confidence level in the automated decision at $\emptyset(x) = 0.997$, which resides within the unacceptable range of $0.80 < \emptyset(x) \leq 1$

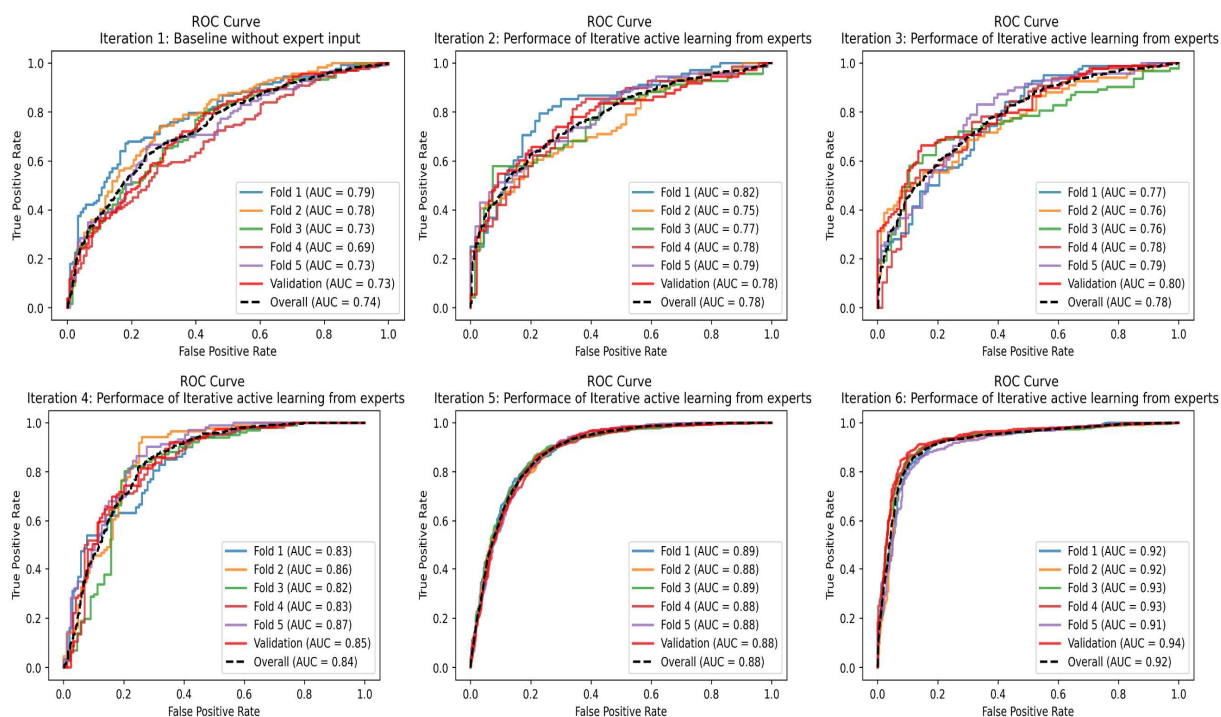

**Fig 7.** AUC performance of six iterations on the dataset retraining with the annotated samples

## 4. Conclusion

This study presents the DeFi TrustBoost Framework, which combines blockchain technology and Explainable AI to address the challenges faced by lenders supporting low-wealth households accessing funds for a small business. Small business ownership is a critical pathway for these households to increase income. It presented the strategy for storing on-chain and off-chain data in compliance with the data protection laws to audit the automated AI decisions and mitigate potential adversarial attacks. It demonstrated the non-tampering nature of visual explainability for individual decisions, which enhances the system's performance over time by refinement system with the low-confident samples annotated by experts. This framework can improve reliability and trust in AI and manual decisions, benefiting underserved individuals seeking funding and other decentralized finance applications.

**Acknowledgment**

We extend our gratitude to Dr. Richard Buchinger from Chain Crunch Labs and J.P. Morgan Chase for their technical support. We also thank Open Trellis for their valuable knowledge exchange on small business loans and ongoing support on impact case studies.